\documentclass[fleqn,usenatbib,useAMS]{mnras}

\usepackage{graphicx}   % Including figure files
\usepackage{amsmath}    % Advanced maths commands
\usepackage{amssymb}    % Extra maths symbols
\usepackage{multicol}        % Multi-column entries in tables
\usepackage{bm}         % Bold maths symbols, including upright Greek
\usepackage{pdflscape}  % Landscape pages

\newcommand{\hmsun}{h^{-1}{\rm M}_\odot}
\newcommand{\hmpc}{h^{-1}{\rm Mpc}}
\newcommand{\kms}{{\rm ~km/s}}

\usepackage[T1]{fontenc}
\usepackage{ae,aecompl}

\usepackage{txfonts}
\title[Into the void]{Structure and dynamics in low density regions: galaxy-galaxy correlations inside cosmic voids}

\author[Andr\'es N. Ruiz et al.]{\parbox[t]{\textwidth}{\vspace{-1cm}
        Andr\'es N. Ruiz$^{1,2}$\thanks{E-mail: andres.ruiz@unc.edu.ar}, 
        Ignacio G. Alfaro$^{1}$ $\&$ 
        Diego Garcia Lambas$^{1,2}$} \\
$^{1}$ Instituto de Astronom\'{\i}a Te\'orica y Experimental (CONICET-UNC), Laprida 854, X5000BGR, C\'ordoba, Argentina\\
$^{2}$ Observatorio Astron\'omico, Universidad Nacional de C\'ordoba, Laprida 854, X5000BGR, C\'ordoba, Argentina
}

\date{Accepted XXX. Received YYY; in original form ZZZ}

\pagerange{\pageref{firstpage}--\pageref{lastpage}} \pubyear{2017}

\def\LaTeX{L\kern-.36em\raise.3ex\hbox{a}\kern-.15em
    T\kern-.1667em\lower.7ex\hbox{E}\kern-.125emX}

\begin{document}
\label{firstpage}
\maketitle

\begin{abstract} %{{{

We compute the galaxy-galaxy correlation function of low-luminosity SDSS-DR7
galaxies $(-20 < M_{\rm r} - 5\log_{10}(h) < -18)$ inside cosmic voids
identified in a volume limited sample of  galaxies at $z=0.085$. 
To identify voids, we use bright galaxies with $M_{\rm r} - 5\log_{10}(h) <
-20.0$. We find that structure in voids as traced by faint galaxies is
mildly non-linear as compared with the general population of galaxies
with similar luminosities. This implies a redshift-space correlation 
function with a similar shape than the real-space correlation albeit a
normalization factor. 
The redshift space distortions of void galaxies allow to calculate pairwise 
velocity distributions which are consistent with an exponential model with 
a pairwise velocity dispersion of $w \sim 50-70 \kms$, significantly lower 
than the global value of $w \sim 500\kms$. 
We also find that the internal structure of voids as traced by faint galaxies 
is independent of void environment, namely the correlation functions of galaxies 
residing in void-in-void or void-in-shell regions are identical within uncertainties.
We have tested all our results with the semi-analytic catalogue MDPL2-\textsc{Sag} 
finding a suitable agreement with the observations in all topics studied.

\end{abstract} %}}}


%%%%%%%%%%%%%%%%%%%%%%%%%%%%%%%%%%%%%%%%%%%%%%%%%%%%%%%%%%%%%%%%%%%%%%%%%%%%%%%%%%%%%%%%%%%%%
%%%%%%%%%%%%%%%%%%%%%%%%%%%%%%%%%%%%%%%%%%%%%%%%%%%%%%%%%%%%%%%%%%%%%%%%%%%%%%%%%%%%%%%%%%%%%
\section{Introduction}

The large-scale structure of the Universe can be understood as a complex net of
filaments and plane structures of dark matter which intersect forming massive 
virialized systems, regions of preferred galaxy formation.
As a result, between these structures, large-scale underdense regions emerge, known as
cosmic voids.
Depending on the identification algorithm, several properties such  as topology and
fraction of mass or galaxies inside the voids can significantly vary
\citep{colberg_voids_2008,cautum_voids_2018}.
One of the simplest way to define cosmic voids is to considerer them as expanding
spherical regions with 10 to 20 percent of the mean density of the Universe
\citep{padilla_voids_2005,ceccarelli_voids_2006,ruiz_clues_2015}.

Observationally, galaxies residing cosmic voids have been studied from different 
point of views and in several galaxy catalogues. 
By construction, voids lack a significant fraction of luminous galaxies, and so their 
inner structure can be traced by faint objects \citep{lindner_voidgxs_1996,alpaslan_voidgxs_2014}.
It should be noticed that, induced by void expansion, galaxies residing voids can 
perceive a local Hubble constant larger than the average \citep{tomita_model_2000}.
Both the large local Hubble constant and a the low mass density conspire against 
structure growth which is expected to be significantly lower than elsewhere, 
providing the galaxies that populate voids very different astrophysical and dynamical 
properties compare to those in higher density regions. 
Regarding to their photometric properties, void galaxies tend to be fainter, bluer 
and late-type \citep{rojas_voidgxs_2004,hoyle_voidgxs_2005,patiri_voidgxs_2006,
ceccarelli_voidgxs_2008, hoyle_voidgxs_2012}. 
Also, the spectroscopic properties of galaxies within voids show younger stellar
populations and higher star formation rates \citep{rojas_voidgxs_2005}.
All these properties depend on the emptiness of the host void as well
\citep{tavasoli_voidgxs_2015}.

The inner structure of cosmic voids has also been studied with detail in numerical
simulations by \citet{gottlober_void_2003} and \citet{aragoncalvo_voids_2013}, 
unveiling the complex structure traced by galaxies in these underdense regions.
In this paper we address the two-point correlation function and derive pairwise 
velocitiy distributions of faint galaxies within voids, both in observations and in 
a cosmological simulation with a semi-analytic model implementation. 

It should be noticed that faint galaxies in rich clusters type environments behave
in a similar fashion than luminous ones since the dynamics is globally dominated 
by the cluster potential. 
By contrast, inside voids, no rich clusters are expected, so that the dynamics of faint
galaxies may reflect more faithfully initial conditions unaffected by induced strongly 
non-linear peculiar motions.

This paper is organized as follows. In Sec. \ref{sec:data} we describe the
observed and simulated galaxy catalogues used. 
In Sec. \ref{sec:xi} we present the results of redshift and real-space inferred
two point correlation functions for galaxies inside voids and compare to the
global results. 
In Sec. \ref{sec:xisp_fv} we analyse the corresponding 2D correlation function
$\xi(\sigma,\pi)$ and derive pairwise velocity distributions.
We also compare in Sec. \ref{sec:xiRS}, the possible clustering dependence on
global void environment.
The results obtained are listed in Sec. \ref{sec:conclusion}.

%%%%%%%%%%%%%%%%%%%%%%%%%%%%%%%%%%%%%%%%%%%%%%%%%%%%%%%%%%%%%%%%%%%%%%%%%%%%%%%%%%%%%%%%%%%%%
%%%%%%%%%%%%%%%%%%%%%%%%%%%%%%%%%%%%%%%%%%%%%%%%%%%%%%%%%%%%%%%%%%%%%%%%%%%%%%%%%%%%%%%%%%%%%
\section{Data}
\label{sec:data}

In this section we present both the observational and the simulated galaxy
catalogues used in this work. Also, we describe the void identification procedure and 
the derived void catalogues.

\subsection{Observed galaxies}

We use  the Main Galaxy Sample of the Sloan Digital Sky Survey Data Release 7
\citep[SDSS-DR7,][]{york_sdss_2000,strauss_sdss_2002,abazajian_dr7_2009}, which
comprises nearly a millon galaxies up to $z \sim 0.3$ with an upper limiting apparent
magnitude in the $r$-band $m_r^{\rm lim} =  17.77$.
From this catalogue, we select all galaxies with $r$-band magnitudes $14.5 \le
m_{\rm r} \le 17.77$ in the redshift range $0.02 \le z \le 0.085$, comprising a total of
approximately $2.5 \times 10^5$ galaxies.

\subsection{Simulated galaxies}

The simulated semianalytic galaxies used in our analysis were extracted from the public
release \textsc{MultiDark-Galaxies} \citep{knebe_multidark_2018}, available at
the \textsc{CosmoSim} database\footnote{hhtp://www.cosmosim.org}. 
Specifically, we use the catalogue
MDPL2-\textsc{Sag}\footnote{doi:10.17876/cosmosim/mdpl2/007}, which is derived
by populating the dark matter haloes of the MDPL2 simulation using the
semi-analytic model of galaxy formation and evolution \textsc{Sag} 
\citep{cora_sag_2018}.

The MDPL2 simulation belongs to the \textsc{MultiDark} suite of simulations
\citep{riebe_multidark_2013, klypin_mdpl2_2016} and is also available at the
\textsc{CosmoSim} database.
This simulation follows the evolution of $3840^3$ dark matter particles in a
cubic comoving box of $1000\hmpc$ on a side. The adopted cosmology is a flat
$\Lambda$CDM with cosmological parameters: $\Omega_{\rm m} = 0.307$,
$\Omega_{\rm \Lambda}=0.693$, $\Omega_{\rm b}=0.048$, $\sigma_8 = 0.823$,
$n=0.96$ and $h=0.678$, consistent with \textsc{Planck} results
\citep{planck_cosmology_2014,planck_cosmology_2016}.
The dark matter haloes have been identified with \textsc{Rockstar}
\citep{behroozi_rockstar_2013}, finding $\sim 127 \times 10^6$ haloes with at
least 20 particles, and the merger trees were constructed using
\textsc{ConsistentTrees} \citep{behroozi_trees_2013}.

The \textsc{Sag} model includes most of the relevant physical processes in 
galaxy formation and evolution, such as radiative cooling of hot gas, 
star formation, feedback from supernova explosions, chemical enrichment, 
growth of supermassive black holes, AGN feedback and starburst via disc 
instabilities and galaxy mergers.
This model was calibrated to generate a galaxy catalogue using the MDPL2
simulation, which together with the catalogues constructed using \textsc{Sage}
\citep{croton_sage_2016} and \textsc{Galacticus} \citep{benson_galacticus_2012}, 
conform the \textsc{MultiDark-Galaxies} database previously mentioned.

The complete MDPL2-\textsc{Sag} catalogues has $\sim 194 \times 10^6$
galaxies at $z=0$, although we keep only those galaxies with $r$-band absolute 
magnitudes in the range $M_{\rm r} - 5\log_{10}(h) \le -16.0$ and stellar masses 
$M_\star \ge 10^8 \hmsun$, nearly $91 \times 10^6$ simulated galaxies.

\subsection{Void catalogues and void galaxies}

In order to identify voids both in the SDSS-DR7 and MDPL2-\textsc{Sag} catalogues, we
use the algorithm described in \citet{ruiz_clues_2015}.
Briefly, the algorithm starts with a Voronoi tessellation \citep{voronoi_1908} 
of the density field, using  galaxies as tracers.
Each Voronoi cell has an associated density which is given by the inverse of the cell 
volume, $\rho_{\rm cell} = 1/V_{\rm cell}$.
Defining the density contrast as 
\begin{equation}
\delta = \frac{\rho_{\rm cell}}{\bar{\rho}} - 1,
\end{equation}
where $\bar{\rho}$ is the mean density of tracers, we select any Voronoi cell 
with $\delta < -0.8$ as a center of an underdense region.
From theses centers, we select as void candidates all spherical volumes with an
integrated density contrast which satisfy
\begin{equation}
\Delta(R_{\rm void}) = \frac{3}{R^3_{\rm void}} \int_0^{R_{\rm void}} 
\delta(r) r^2 dr < -0.9,
\end{equation}
where $R_{\rm void}$ is the void radius. 
Once these void candidates are identified, we repeat the computation of
$\Delta$ in a randomly displaced center around the previous one, where these
are only accepted if the new $R_{\rm void}$ is larger than the
previous one. 
This procedure is repeated several times, mimicking a random walk around the
original center, alloing to obtain void candidates with centers located in the true
local minima of the density field.
Finally, the void catalogue is formed by the largest void candidates which
do not overlap with any other void candidate.

In the case of the SDSS-DR7 sample, we select all galaxies with absolute
magnitudes in the $r$-band brighter than $M_{\rm r} - 5\log_{10}(h) = -20.0$
(the limiting magnitude for a complete volume limited sample at $z=0.085$ is
$M_{\rm r} - 5\log_{10}(h) = -19.4$).
This subsample of $\sim 7.3 \times 10^4$ galaxies, which corresponds to a
galaxy volume number density of $n=6.4\times 10^{-3}$, was used as tracers to
identify spherical cosmic voids, obtaining $167$ voids with radii in the range
$7-24 \hmpc$.
Void identification procedure considers a {\sc HEALPix} \citep{healpix_2005}
angular mask of the SDSS-DR7 that takes into account catalogue boundaries and
holes within the survey.
It is important to note that we  only consider those voids which do not include 
any boundary of the catalogue.
Finally, we consider all galaxies in SDSS-DR7 in the redshift range $0.02 \le
z \le 0.085$ and $-18.0 \ge M_{\rm r} - 5\log(h) \ge -20.0$ in order to
populate the voids with low luminosity galaxies. 

For the MDPL2-\textsc{Sag} catalogue, to identify voids comparable
to those identified in the SDSS-DR7 data, we construct a subsample of galaxies
with the same volume density than in the SDSS-DR7 by selecting all galaxies
brighter than $M_{\rm r} - 5\log(h) = -20.6$.
This cut is different than that used in SDSS-DR7 mainly due to the differences 
in the luminosity functions of observations and simulations.
Voids identified in the full MDPL2 box with the same volume density of tracers
have the same radii range than those identified in the SDSS-DR7, allowing us a
proper comparison between both datasets.
In order to take into account redshift-space distortions, the 3D positions of
simulated galaxies were transformed to redshift-space by affecting the
z-coordinate by peculiar velocities, z$_{\rm redshift-space} = {\rm z} + v_{\rm
z}/100h^{-1}$, where $v_{\rm z}$ is the z-component of the peculiar velocity of the
galaxy.
This subsamble comprises $6.4 \times 10^6$ galaxies, and we use it to identify
$17694$ voids with the same radii range than those in the SDSS-DR7. To populate
these voids with low luminosity galaxies, we consider all galaxies in the
catalogue with $-18.6 \ge M_{\rm r} - 5\log(h) \ge -20.6$.

\begin{figure}
	\includegraphics[width=\columnwidth]{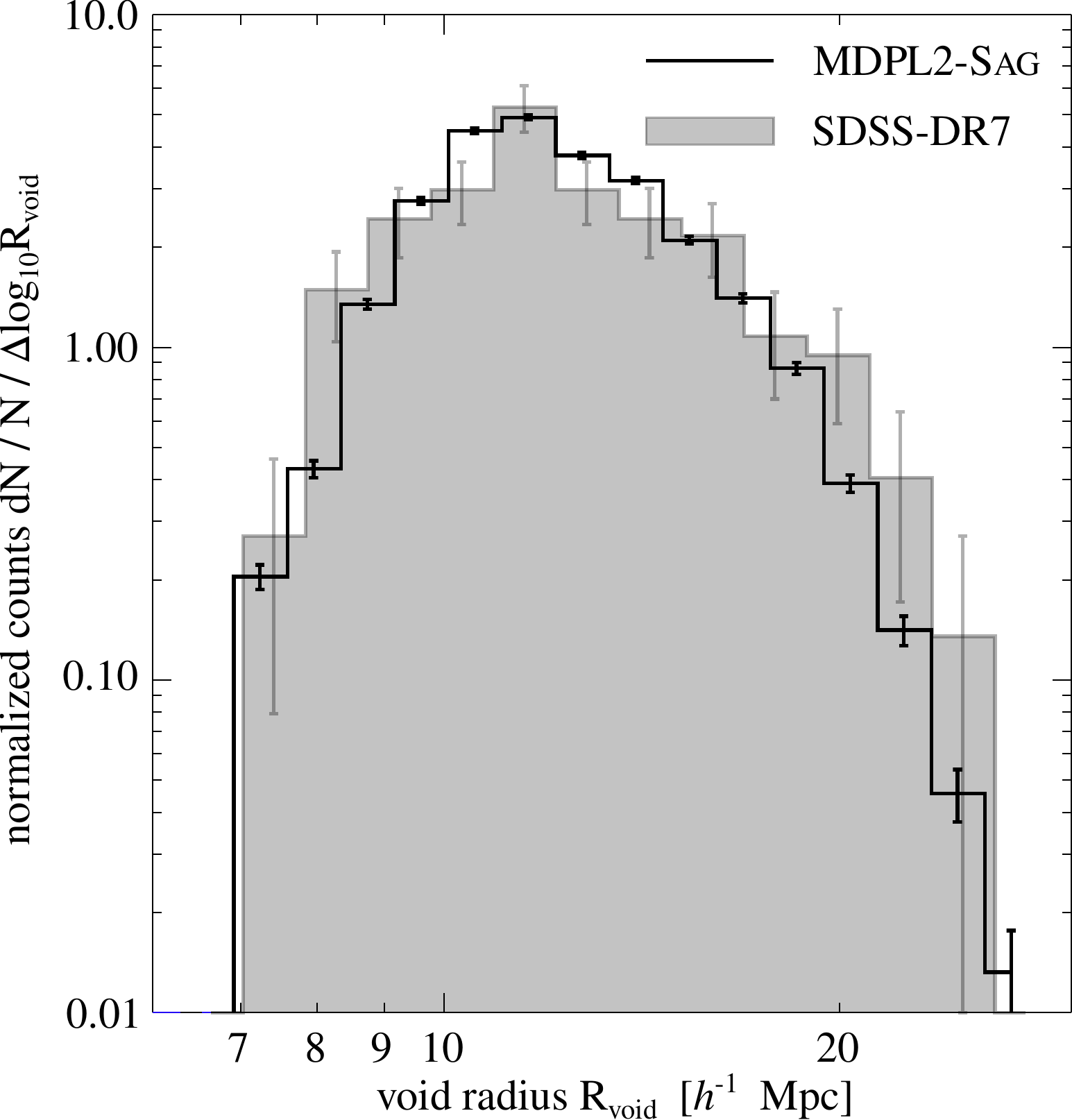}
    \caption{Normalized number counts of voids as a function of void radius. 
    The distribution for MDPL2-\textsc{Sag} voids are shown with black solid line and for 
    SDSS-DR7 voids with the grey shaded area. In both cases, error bars correspond to 
    Poisson uncertainties.}
    \label{fig:hist_rad}
\end{figure}

Fig. \ref{fig:hist_rad} show the normalized void radius distributions from both
simulated and observed catalogues. 
The distribution for voids identified in MDPL2-\textsc{Sag} catalogue is plotted 
with the solid black line, and the one corresponding to SDSS-DR7 voids whit the 
grey shaded histogram. 
Error bars correspond to Poisson uncertainties in both cases.

%%%%%%%%%%%%%%%%%%%%%%%%%%%%%%%%%%%%%%%%%%%%%%%%%%%%%%%%%%%%%%%%%%%%%%%%%%%%%%%%%%%%%%%%%%%%%
%%%%%%%%%%%%%%%%%%%%%%%%%%%%%%%%%%%%%%%%%%%%%%%%%%%%%%%%%%%%%%%%%%%%%%%%%%%%%%%%%%%%%%%%%%%%%
\section{The two-point galaxy-galaxy correlation function} \label{sec:xi}

\subsection{Redshift and real-space correlation functions}

We compute the redshift-space correlation functions, $\xi(s)$, we using the
classical estimator of \citet{davis_xi_1983},
\begin{equation} 
\xi(s) = \frac{DD(s)}{DR(s)}\frac{N_{\rm R}}{N_{\rm D}} - 1,
\end{equation}
where $N_{\rm D}$ is the number of galaxies, $N_{\rm R}$ is the number of
random tracers, $DD$ is the number of galaxy-galaxy pairs and $DR$ is the
number of galaxy-random tracer pairs.
We have also tested other estimators finding similar results. .

For both galaxy catalogues (observed and simulated), we compute $\xi(s)$ for two
samples: (i) galaxies inside cosmic voids, namely \textit{void galaxies}, and
(ii) galaxies in the full catalogue, namely \textit{all galaxies}.
For sample (i) we stack all pairs of galaxies in all voids into one measure of
$\xi(s)$. We use random distributions of tracers in spheres centered in voids
with the same radii, so we take into account border effects in the computation of
$DR$ pairs consistent with the distribution of galaxies constrained to reside in
cosmic voids.
We also compute $\xi(s)$ for sample (ii) where the galaxies have the same magnitude
distribution of void galaxies,  allowing for a fair comparison between these
two equal-luminosity galaxy samples. 

In the case of SDSS-DR7 catalogue, the unclustered random distribution of tracers is 
generated using the method presented in \citet{cole_random_2011} within the SDSS-DR7 mask.
In order to compute $\xi(s)$ for void galaxies we use a distribution of random
tracers with spherical distributions that fill the void volumes.
For MDPL2-\textsc{Sag} catalogue, the number of $DR$ pairs for all galaxies can be 
computed analytically, whereas for void galaxies we use random 
distributions inside spheres of the same distribution of void radii and with the same number density of tracers. 

The real-space correlation function, $\xi(r)$, for SDSS-DR7 galaxies was derived 
from the redshift-space counterpart by applying the inversion method presented 
by \citet{saunders_invertion_1992}. 
In this method, the real-space correlation function can be computed
using the projected correlation function $\Xi(\sigma)$ (see Sec \ref{sec:model_fv})
through 
\begin{equation}
	\xi(r_i) = -\frac{1}{\bar{\pi}} \sum_{j \ge i}
	\frac{\Xi_{j+1}-\Xi_j}{\sigma_{j+1}-\sigma_j} \ln \left(
	\frac{\sigma_{j+1} + \sqrt{\sigma_{j+1}^2 + \sigma_i^2}} {\sigma_j +
	\sqrt{\sigma_j^2 - \sigma_i^2}} \right) \label{eq:invertion}
\end{equation}
where $r_i = \sigma_i$, $\Xi_i = \Xi(\sigma_i)$ and $\bar{\pi} = 3.14159...$,
to distinguish it from $\pi$, the line-of-sight component of the galaxy-galaxy
distance we will introduce in Sec. \ref{sec:xisp_fv}.
For MPDL2-\textsc{Sag} galaxies, $\xi(r)$ is directly measured from the
simulated catalogue. 

%%%%%%%%%%%%%%%%%%%%%%%%%%%%%%%%%%%%%%%%%%%%%%%%%%%%%%%%%%%%%%%%%%%%%%%%%%%%%%%%%%%%%%%%%%%%%
\subsection {Results for SDSS-DR7 and MDPL2-\textsc{Sag} galaxies}
\label{sec:xi_results}

\begin{figure}
	\includegraphics[width=\columnwidth]{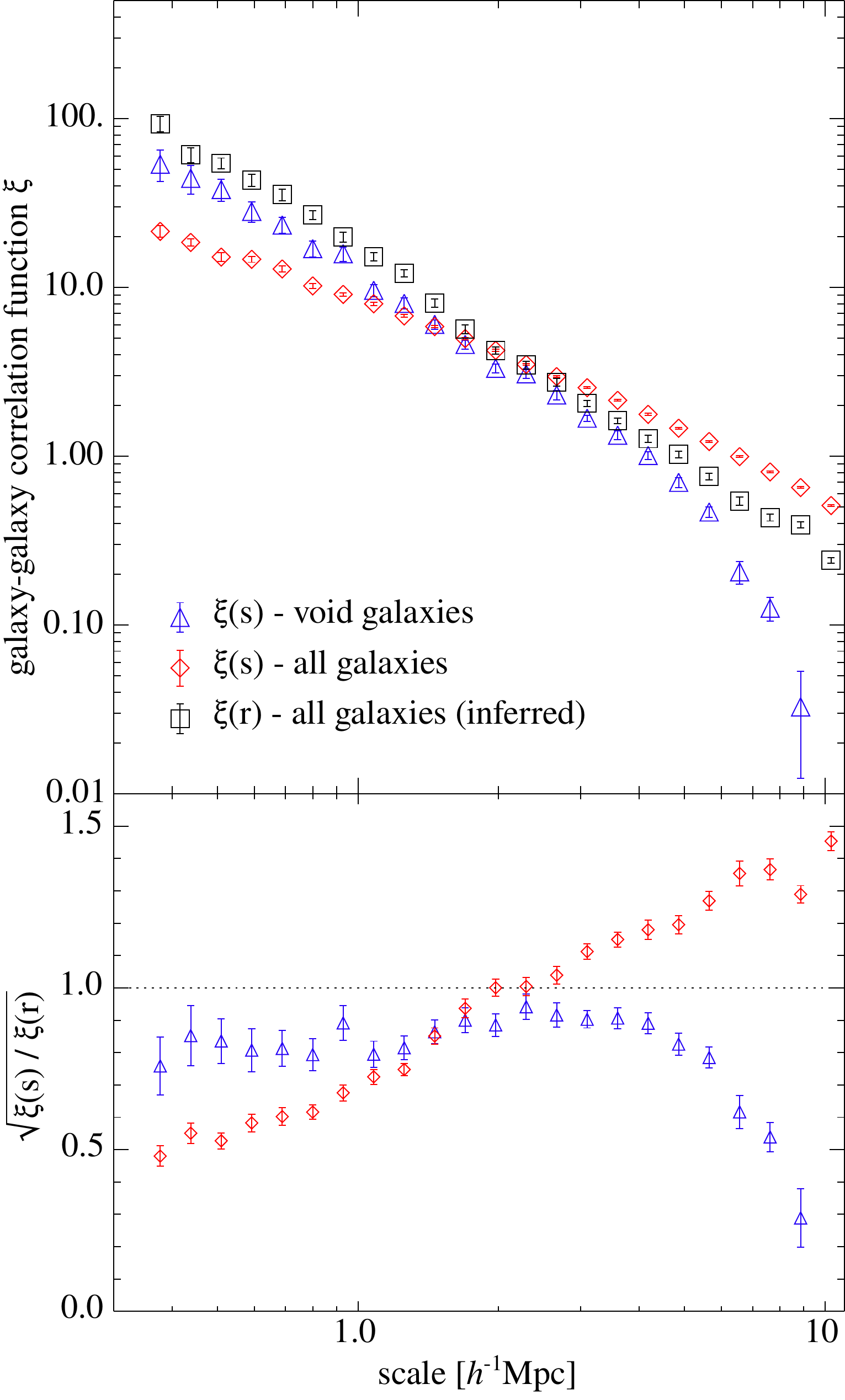}
    \caption{\textit{Top panel:} Galaxy-galaxy correlation function in 
    the SDSS-DR7 catalogue. Blue triangles represent the redshift-space 
    correlation functions $\xi(s)$ for galaxies inside voids stacked for all
    the voids, meanwhile the red diamonds show the $\xi(s)$ for all 
    galaxies in the catalogue with the same luminosity distribution than 
    galaxies inside voids. Black squares show the real-space $\xi(r)$ correlation 
    function for all galaxies obtained via the inversion of $\Xi(\sigma)$ 
    \citep{saunders_invertion_1992}. Errors represent the variance 
    in galaxy pairs estimated using Jackknife resampling. 
    \textit{Bottom panel:} ratio between redshift and real-space correlations
    defined as $\sqrt{\xi(s)/\xi(r)}$. Errors were obtained with
    error propagation.}
    \label{fig:xi_sdss}
\end{figure}

\begin{figure}
	\includegraphics[width=\columnwidth]{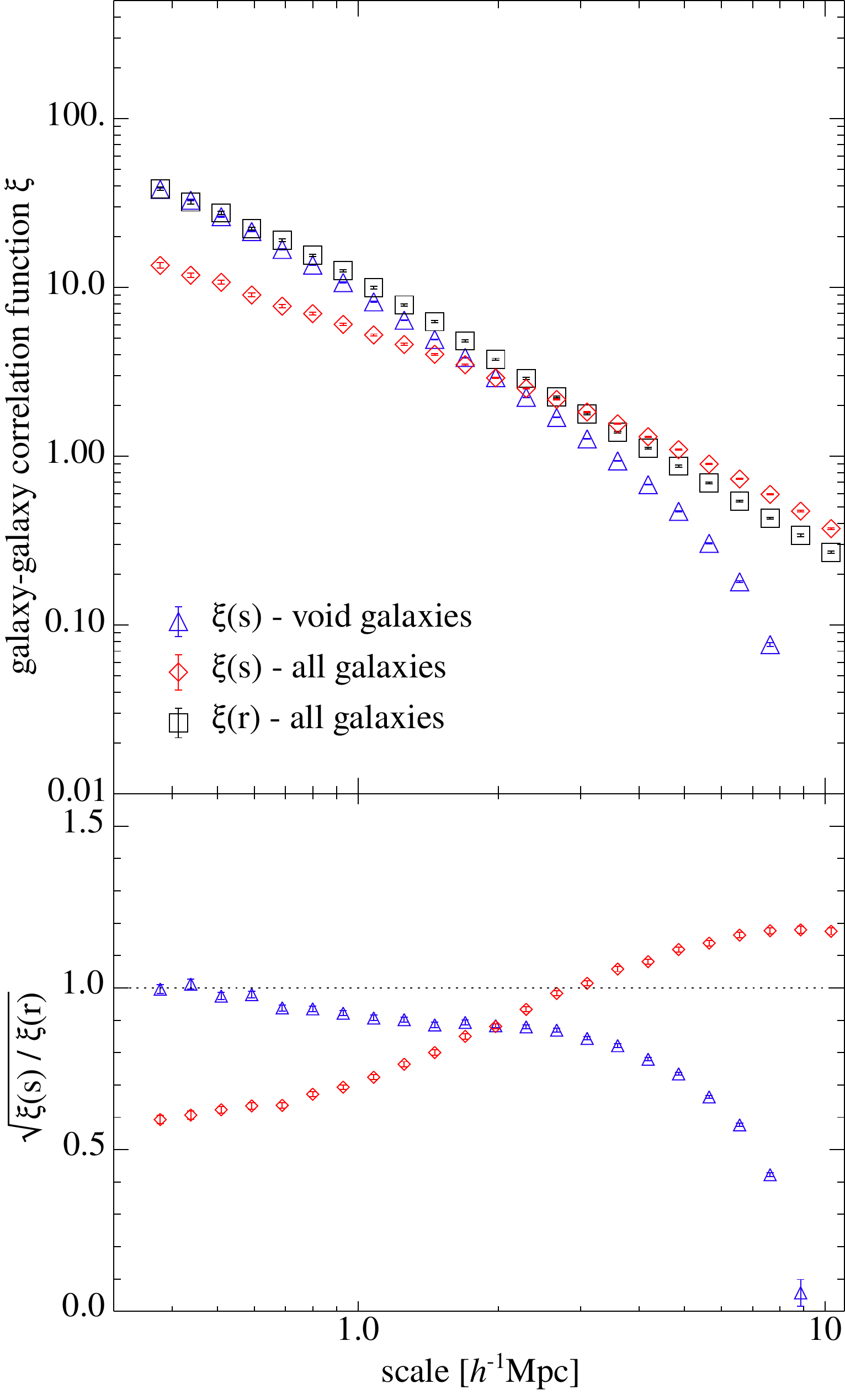}
    \caption{Same as Fig. \ref{fig:xi_sdss} but for the MDPL2-\textsc{Sag}
    catalogue. The only difference is that in this case $\xi(r)$ can 
    be computed directly from the simulated galaxies unaffected by peculiar 
    velocities distortions.}
    \label{fig:xi_mdpl}
\end{figure}

In Fig. \ref{fig:xi_sdss} we show the real-space $\xi(r)$ and redshift-space
$\xi(s)$ two point galaxy correlation functions computed for SDSS-DR7.
In the upper panel are given the resulting $\xi(s)$ for all galaxies (red diamonds)
and for galaxies in voids (blue triangles). 
The black squares correspond to $\xi(r)$ for all galaxies inferred via 
Eq.~(\ref{eq:invertion}).
In all cases, the errorbars represent the uncertainties estimated via Jackknife
resampling \citep{efron_jackknife_1982} in galaxy pairs.
As it can be seen in this panel, at small scales ($r < 5\hmpc$) the shape 
of the real-space  correlation function of the  global population differs  
significantly from the corresponding redshift-space function while on the 
contrary, galaxies in cosmic voids have redshift-space correlation function 
remarkably close to the real-space measures.
This is  more clearly shown  in the bottom panel of this figure where the 
ratio between redshift and real-space correlation functions, defined as $\sqrt{\xi(s)/\xi(r)}$, 
for void galaxies is nearly constant at $\sim$ $0.8$ in the range $0.3 - 3.0  \hmpc$.  
By contrast, in this same range of separations, the ratio for the global
population changes by a factor larger than $2$.

The corresponding results for the semi-analytic model MDPL2-\textsc{Sag} are
shown in Fig. \ref{fig:xi_mdpl}.
Again, $\xi(s)$ for all galaxies is shown in red diamonds, and for galaxies in 
voids, in blue triangles.
The resulting $\xi(r)$ for all galaxies, shown with black squares, is computed
directly form the MDPL2-\textsc{Sag} galaxies unaffected by peculiar velocities.
As it can be seen here, the results are remarkably similar although with lower
errorbars due to the significantly larger number of semi-analytical galaxies
and voids.
In the case of MDPL2-\textsc{Sag}, \textbf{the ratio $\sqrt{\xi(s)/\xi(r)}$} is also 
approximately constant in the range $0.3 - 3.0 \hmpc$, while the global one also 
changes by a factor $\sim$ 2.

As can be seen in Figs. \ref{fig:xi_sdss} and \ref{fig:xi_mdpl}, $\xi(s)$ for
void galaxies depart from a power-law behavior for scales $s > 3.0\hmpc$. 
This in consistent with the absence of a significant two-halo term in the
galaxy-galaxy correlations inside voids due to the lack of massive haloes associated
to bright galaxies.
Given the void radii distributions (see Fig. \ref{fig:hist_rad}), galaxy-galaxy 
correlations are well determined up to scales of $\sim 10\hmpc$, the maximum scale 
shown in these figures.

It is also important to note in Fig. \ref{fig:xi_mdpl} the lack of power of 
$\xi(r)$ for $r<1\hmpc$, a feature that can be understood due to orphan galaxies 
in semi-analytical models. 
Orphan galaxies are galaxies which can not retain their host subhaloe, a
phenomenon that can occur by physical mechanism such as tidal stripping, or just by
resolution limitations in the subhaloe identification
\citep{onions_subhalo_2012}.
The treatment of this particular type of galaxies differs from model to model
\citep{knebe_sam_2015}, having an impact in the galaxy-galaxy correlation function
at small scales \citep{pujol_xisam_2017,knebe_multidark_2018}.
The \textsc{Sag} model used in this work considers the orphan galaxies,
deriving their positions and velocities from an orbital integration.
This treatment allows to obtain an adequate radial distribution of satellite
galaxies, however, it does not trace faithfully the spatial distribution of the
true population of faint galaxies, a fact that is expected to be particularly
serious in high density regions where nonlinearities dominate the dynamics. 
Inside cosmic voids, however, these effects are expected to be less important
given that the dynamical behaviour of galaxies in regions lacking strong
mass concentrations is expected to be only  mildly non-linear.   

\begin{figure}
	\includegraphics[width=\columnwidth]{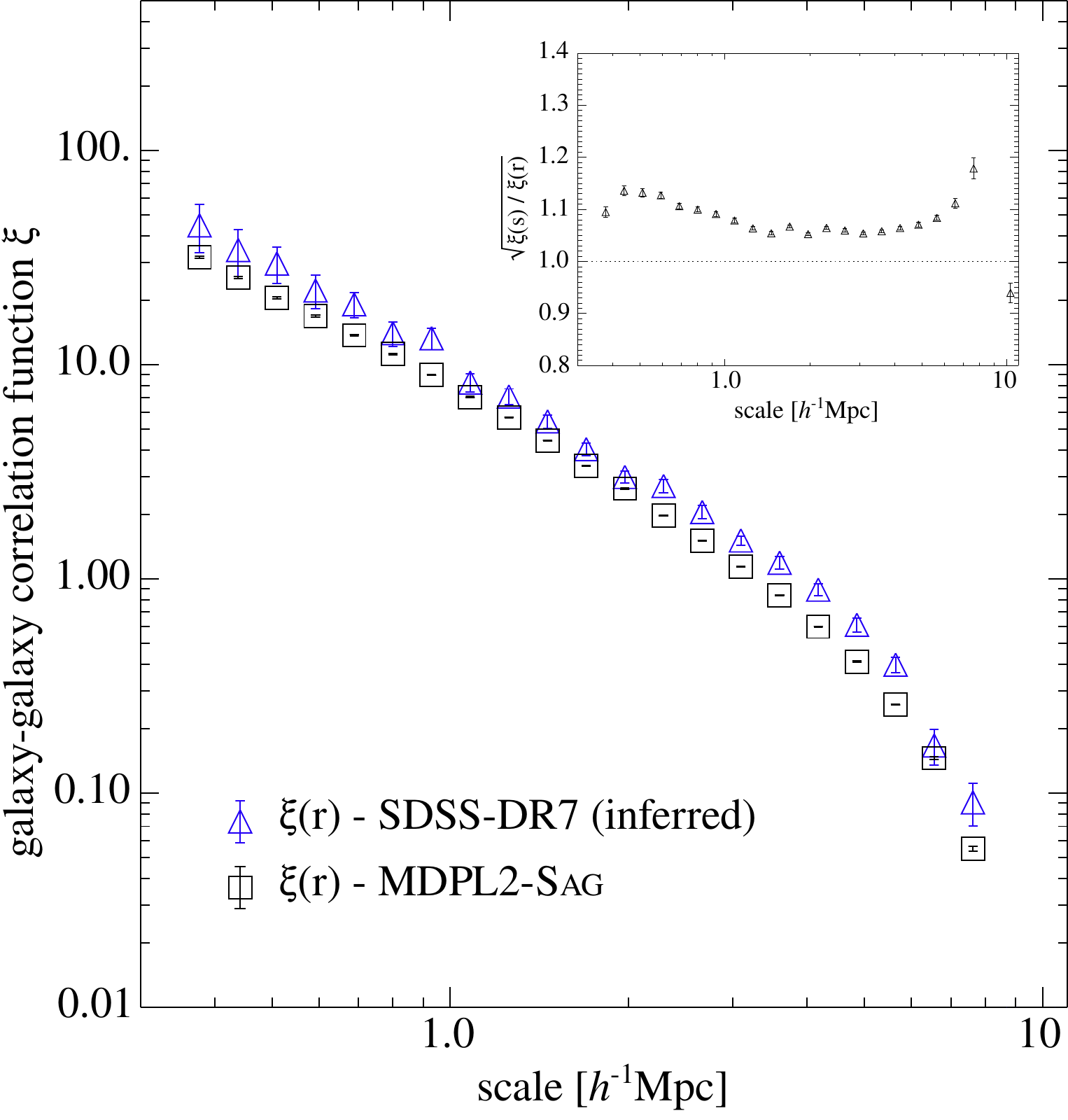}
    \caption{Galaxy-galaxy correlation functions of void galaxies in real-space
    for the MDPL2-\textsc{Sag} (black squares) and for the SDSS-DR7 (blue triangles).
    The inset figure shows the \textbf{ratio $\sqrt{\xi(s)/\xi(r)}$} measured for
    MDPL2-\textsc{Sag} void galaxies, which was used to infer $\xi(r)$ for SDSS-DR7
    void galaxies (see text for details). Errorbars correspond to Jackknife estimations
    of the uncertainties.}
    \label{fig:xi_void}
\end{figure}

In Fig. \ref{fig:xi_void} we show the real-space correlation function $\xi(r)$
for MDPL2-\textsc{Sag} (direct measure) with black squares and that derived 
for void galaxies SDSS-DR7 with blue triangles. 
We notice that the real-space $\xi(r)$ for SDSS-DR7 void galaxies obtained by the 
direct inversion of Eq.~(\ref{eq:invertion}) is noisy given the low number 
statistics with respect to the MDPL2-\textsc{Sag} void galaxies.
Therefore, we assume the same ratio between $\xi(r)$ and $\xi(s)$ of the
simulations to infer the spatial correlation function $\xi(r)$ of SDSS-DR7 from
the corresponding redshift-space $\xi(s)$ determination. 
We notice that the inferred $\xi(r)$ by this method is entirely consistent with 
that derived by direct inversion using Eq.~(\ref{eq:invertion}), albeit much 
smoother.
This ratio is shown in the inset of Fig. \ref{fig:xi_void}.
By inspection to  this figure it can be seen a very suitable agreement of the
two measures showing that, inside voids, the distribution of
MDPL2-\textsc{Sag} faint galaxies is in agreement with that  observed in the
SDSS-DR7 data.

\begin{table}
\centering
\begin{tabular}{ccccc}
          & MDPL2-\textsc{Sag} & MDPL2-\textsc{Sag} & SDSS-DR7        & SDSS-DR7         \\
          & all galaxies       & void galaxies      & all galaxies    & void galaxies    \\
          \hline
 $r_0$    & $4.5  \pm 0.1$     & $3.3 \pm 0.1$      & $4.8  \pm 0.1$  & $3.9  \pm 0.1$    \\
 $\gamma$ & $1.60 \pm 0.02$    & $1.66 \pm 0.03$    & $1.78 \pm 0.02$ & $1.65 \pm 0.03$   \\
\end{tabular}
\caption{Autocorrelation function $\xi(r)$ power-law fitting parameters. For galaxies in 
voids, we restrict the fit to measurements in the range $0.3 - 3.0\hmpc$.}
\label{tab:xi3d}
\end{table}

In Tab. \ref{tab:xi3d} we show the fitting parameters obtained by modeling the
real-space correlation function with a power-law of the form 
\begin{equation} 
\xi(r) = \left(\frac{r}{r_0}\right)^{-\gamma} \label{eq:xir}
\end{equation}
We adopt the range $0.3 - 10 \hmpc$ for all galaxies, and use a more restricted
range of $0.3 - 3.0\hmpc$ for galaxies in voids to avoid border effects.
The purpose of these fits are mainly to provide simple theoretical models for the
2D correlation function $\xi(\sigma,\pi)$ and the derivation of the pairwise velocity
distribution $f(v)$ presented in the next section.

%%%%%%%%%%%%%%%%%%%%%%%%%%%%%%%%%%%%%%%%%%%%%%%%%%%%%%%%%%%%%%%%%%%%%%%%%%%%%%%%%%%%%%%%%%%%%
%%%%%%%%%%%%%%%%%%%%%%%%%%%%%%%%%%%%%%%%%%%%%%%%%%%%%%%%%%%%%%%%%%%%%%%%%%%%%%%%%%%%%%%%%%%%%
\begin{figure*}
    \includegraphics[width=\textwidth]{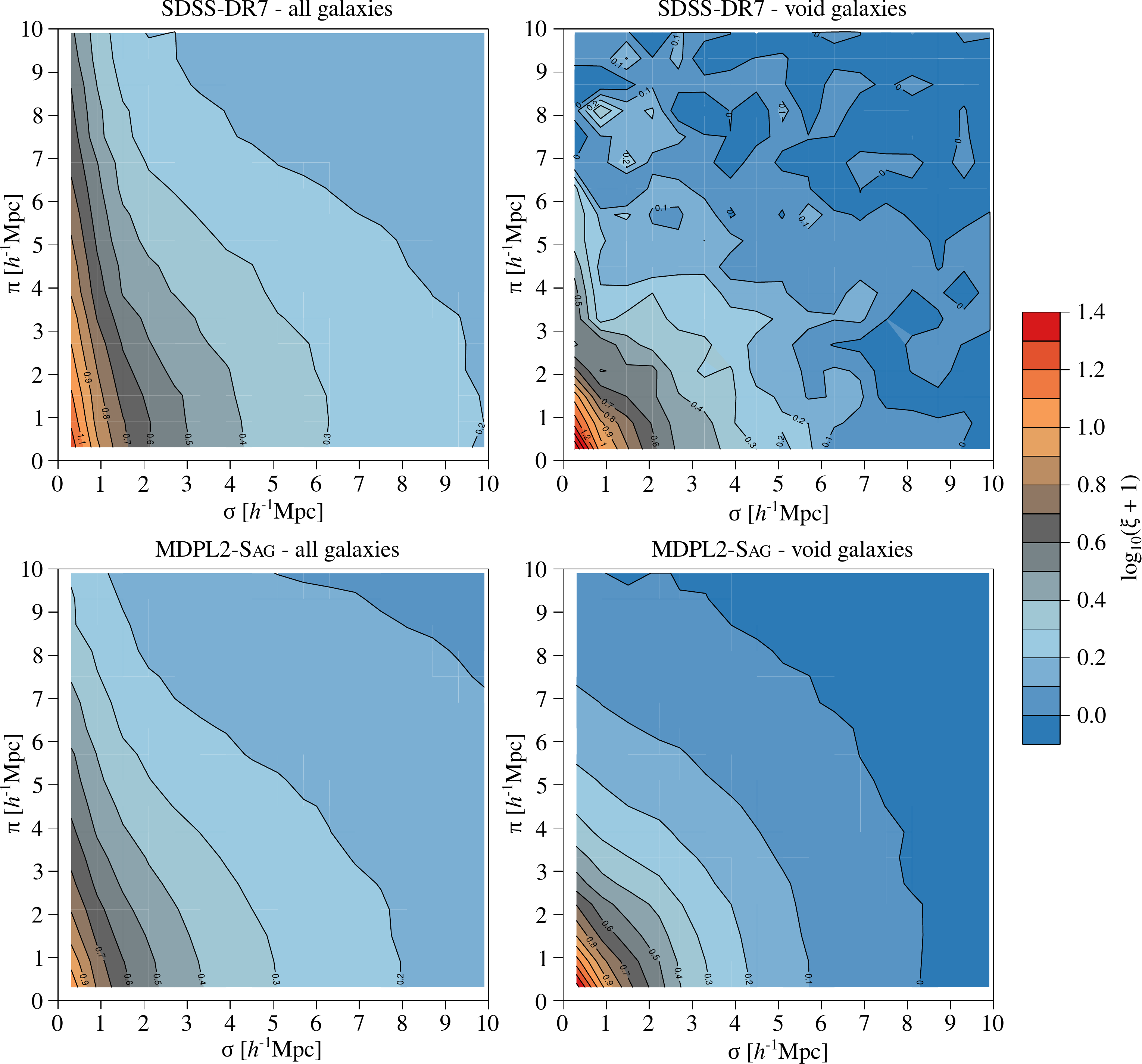}
    \caption{2D galaxy-galaxy correlation function $\xi(\sigma,\pi)$ for SDSS-DR7
    (upper panels) and MDPL2-\textsc{Sag} (bottom panels) galaxies. For both catalogues,
    the results for all galaxies are shown in left panels, meanwhile in the right panels
    are the results for void galaxies. Contours show the different values of 
    $\log_{10}(\xi + 1)$ as indicated in the legends and colour bar.}
    \label{fig:xi_sigma_pi}
\end{figure*}

\section{$\xi(\sigma,\pi)$ and derivation of pairwise velocity distributions}
\label{sec:xisp_fv}

We have also computed the 2D galaxy-galaxy correlation function in the form
$\xi(\sigma,\pi)$, where $\sigma$ and $\pi$ are the projected and line-of-sight
components of the galaxy-galaxy pair distance, respectively.

In Fig. \ref{fig:xi_sigma_pi} we show the $\xi(\sigma,\pi)$ correlation functions 
computed for the SDSS-DR7 (upper panels) and MDPL2-\textsc{Sag} (bottom panels). 
In both cases, left panels correspond to $\xi(\sigma,\pi)$ measured in all
galaxies and right panels to the measurements for void galaxies.
By inspection to these figures it can be seen a general agreement between
simulated and observational data.
As seen in Figs. \ref{fig:xi_sdss} and \ref{fig:xi_mdpl}, galaxies within voids
present much less distortions due to Finger-of-God effect than in the global
population.

The information contained in these correlation functions can be used to
estimate galaxy pairwise velocity distributions, $f(v)$, to have more insight
in the dynamical properties of galaxies inside cosmic voids. 
In the next subsections we describe the procedures to measure and estimate
$f(v)$ from $\xi(\sigma,\pi)$.

%%%%%%%%%%%%%%%%%%%%%%%%%%%%%%%%%%%%%%%%%%%%%%%%%%%%%%%%%%%%%%%%%%%%%%%%%%%%%%%%%%%%%%%%%%
\subsection{Measuring $f(v)$}

In order to determine the pairwise velocity distributions of both SDSS-DR7 and
MDPL2-\textsc{Sag} catalogues, we use the projections of the correlation
functions, $\Xi$, onto the $\sigma$ and $\pi$ axes
\begin{eqnarray} 
\Xi_\sigma = \Xi(\sigma) = 2 \int_0^{\pi_{\rm lim}} \xi(\sigma,\pi) d\pi \label{eq:xi_sigma} \\ 
\Xi_\pi = \Xi(\pi) = 2 \int_0^{\sigma_{ \rm lim}} \xi(\sigma,\pi) d\sigma 
\end{eqnarray}
where $\sigma_{\rm lim} = \pi_{\rm lim} = 20 \hmpc$. The projected correlation
function $\Xi_\sigma$ is not affected by redshift space distortions induced by
pairwise velocities, meanwhile the projection $\Xi_\pi$ arises from a
convolution with the pairwise velocity distribution.
We take the Fourier transform $(\mathcal{F})$ of $\Xi_\sigma$ and $\Xi_\pi$,
and take advantage of the fact than a convolution is just a multiplication in
Fourier-space to recover the pairwise velocity distribution: 
\begin{equation} f(v) = \mathcal{F}^{-1}
\left[\frac{\mathcal{F}[\Xi_\pi]}{\mathcal{F}[\Xi_\sigma]} \right]
\label{eq:fv} \end{equation}

%%%%%%%%%%%%%%%%%%%%%%%%%%%%%%%%%%%%%%%%%%%%%%%%%%%%%%%%%%%%%%%%%%%%%%%%%%%%%%%%%%%%%%%%%%
\subsection{Modeling $f(v)$}
\label{sec:model_fv}

To model the velocity distributions $f(v)$ and obtain an estimation for the
mean pairwise velocity dispersion $w=\langle \sigma_v^2 \rangle^{1/2}$, we 
follow the method described in \citet{hawkins_2df_2003}.

We start with a model of $\xi_{\rm K}(\sigma,\pi)$ which takes into account the
coherent infall velocities \citep{kaiser_infall_1987,hamilton_infall_1992}
\begin{equation} \xi_{\rm K}(\sigma,\pi) = \sum_{i=0,2,4} \xi_i(s)P_i(\cos{\theta})
\end{equation}
where $P_i(x)$ are the Legendre polynomials and $\theta$ is the angle between
$r$ and $\pi$. If we asume a power law for $\xi(r)$ of the form given by 
Eq.~(\ref{eq:xir}), the relations between $\xi_i(s)$ and $\xi(r)$ are given by 
\begin{eqnarray}
\xi_0(s)  &=& \left(1+\frac{2\beta}{3}+\frac{\beta^2}{5}\right)\xi(r) \\ 
\xi_2(s) &=& \left(\frac{4\beta}{3}+\frac{4\beta^2}{7}\right) \left(\frac{\gamma}{\gamma - 3}\right)\xi(r) \\ 
\xi_4(s) &=& \frac{8\beta^2}{35} \left(\frac{\gamma(2+\gamma)}{(3-\gamma)(5-\gamma)}\right)\xi(r) 
\end{eqnarray}
where 
\begin{equation} \beta = \frac{f(\Omega)}{b} \simeq \frac{\Omega_{\rm
m}^{0.6}}{b} \label{eq:beta} \end{equation}
is the parametrization of the large-scale coherent infall,  $f(\Omega)$ is
the linear growth rate of density fluctuations, and $b$ is the linear bias parameter.

Following \citet{peebles_lss_1980}, we obtain a model for $\xi(\sigma,\pi)$, by 
convolution of  $\xi_{\rm K}(\sigma,\pi)$ with the distribution function of pairwise motions 
$f(v)$ 
\begin{equation}
\xi(\sigma,\pi) = \int_{-\infty}^{+\infty} \xi_{\rm K}(\sigma,\pi-v/H_0) f(v) dv.
\end{equation}
We consider the usual assumtion of an exponential form for the distribution of pairwise motions
\begin{equation}
f(v) = \frac{1}{\sqrt{2}~w}\exp{\left(-\frac{\sqrt{2}~|v|}{w}\right)},
\label{eq:exp}
\end{equation}
The assumed exponential form of $f(v)$ provides a suitable description 
of observational and simulated data, better than other distributions as, for
instance a Gaussian model
\citep{sheth_fv_1996,diaferio_fv_1996,hawkins_2df_2003,loveday_gama_2018}.

As mentioned in Sec. \ref{sec:xi_results}, the correlation functions for void
galaxies, both in redshift and real-space, drop for scales bigger than $\sim 3\hmpc$
(see Figs \ref{fig:xi_sdss}, \ref{fig:xi_mdpl} and \ref{fig:xi_void}), due to the
convolution of different void sizes in the stacked galaxy pair counts.
This means that the power-law fits of Tab. \ref{tab:xi3d} are only valid for
$r<3\hmpc$, however, we use them for scales up to $20\hmpc$ in order to model 
$\xi(\sigma,\pi)$ from where we infer $f(v)$ via Eqs.~(\ref{eq:xi_sigma})-(\ref{eq:fv}).
We have tested the variations of the resulting $f(v)$ using different ranges of $\sigma$
and $\pi$ in the modeled correlation, finding no significant changes. 

%%%%%%%%%%%%%%%%%%%%%%%%%%%%%%%%%%%%%%%%%%%%%%%%%%%%%%%%%%%%%%%%%%%%%%%%%%%%%%%%%%%%%%%%%%
\subsection{Derived and measured $f(v)$}
\label{sec:fit}

\begin{figure*}
	\includegraphics[width=\textwidth]{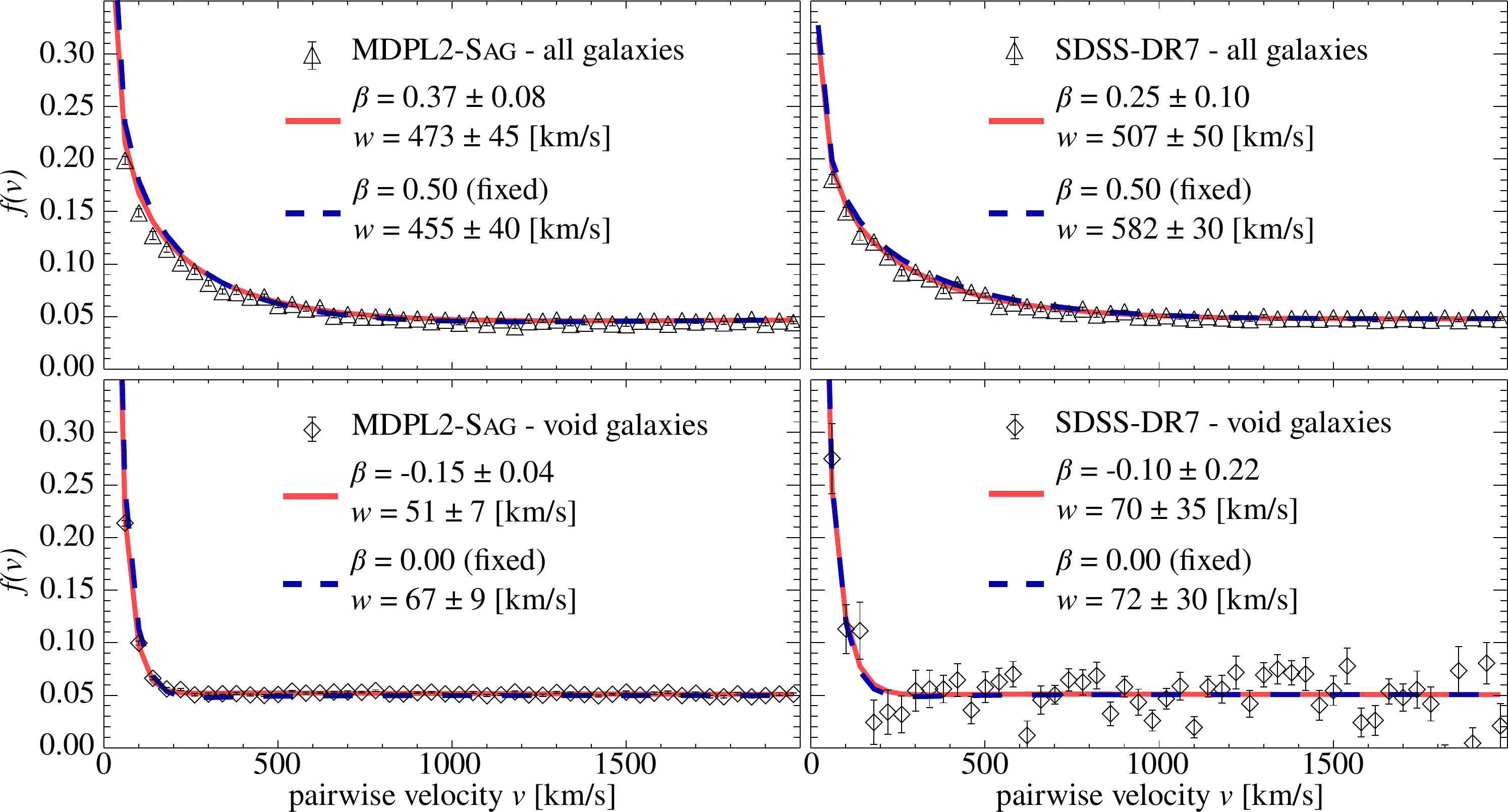}
    \caption{Measured and fitted pairwise velocity distributions. Upper panels  
    correspond to all galaxies for MDPL2-\textsc{Sag} (left) and SDSS-DR7 (right).  
    Bottom panels show the results for void galaxies for 
    MDPL2-\textsc{Sag} (left) and SDSS-DR7 (right). The measured $f(v)$ is 
    represented by triangles (all galaxies sample) and diamonds (void galaxies 
    sample). Two model fits are shown: varying both $\beta$ and $w$ 
    parameters in solid red lines and fixing $\beta$ to a fiducial value and 
    varying only $w$ in dashed blue lines (see text for details).
    The values of the parameters $\beta$ and $w$ are shown in the respective 
    key figures. Errorbars in measured points where estimated via the 
    Jackknife resamplings of $\xi(\sigma,\pi)$.}
    \label{fig:fv}
\end{figure*}

\begin{table}
\centering
\begin{tabular}{ccccc}
            & MDPL2-\textsc{Sag} & MDPL2-\textsc{Sag} & SDSS-DR7        & SDSS-DR7         \\
            & all galaxies       & void galaxies      & all galaxies    & void galaxies    \\
 \hline
 $w$~[km/s] & $473 \pm 45$       & $50 \pm 7$         & $507 \pm 50$    & $70 \pm 35$      \\
 $\beta$    & $0.37 \pm 0.05$    & $-0.15 \pm 0.04$   & $0.25 \pm 0.10$ & $-0.10 \pm 0.22$ \\
 \hline
 $w$~[km/s]      & $455 \pm 40$  & $67 \pm 9$         & $582 \pm 30$    & $72 \pm 30$      \\
 $\beta$ (fixed) & $0.50$        & $0.00$             & $0.50 $         & $0.00$           \\
 \hline
\end{tabular}
\caption{Values of the dynamical parameters $\beta$ and $w$, obtained by fitting
the modeled pairwise velocity distribution to the measurements extracted from 
$\xi(\sigma,\pi)$. The first two rows corresponds to fitting both $\beta$
and $w$, meanwhile in the second two rows are the results obtained by fitting only
$w$ with a fixed $\beta$ value (see text for details).}
\label{tab:fv}
\end{table}

To provide the model fits, both $\beta$ and $w$ are taken as free parameters,
and adjusted to the measured $f(v)$ by minimizing 
\begin{equation} 
\chi^2 =  \sum_{i=1}^{N} \left(\frac{f_{\rm data}(v_i) - f_{\rm model}(v_i)}{\epsilon_i}\right)^2, 
\end{equation}
where $\epsilon_i$ are the data uncertainties and $N$ is the number of velocity
bins.
The values of $f_{\rm model}(v_i)$ are also derived from the modeled
$\xi(\sigma,\pi)$ via Eq.~(\ref{eq:fv}).

The results are given in Tab. \ref{tab:fv}, where it can be noticed the
suitable agreement between MDPL2-\textsc{Sag} and SDSS-DR7 determinations of
the $w$ and $\beta$ parameters. 
Quoted errors in the table correspond to estimates derived by the dispersion of
the fitting parameters in realizations of $f(v)$ data including individual
uncertainties of each velocity bin.

The resulting $f(v)$ distributions and the corresponding fits are shown in Fig.
\ref{fig:fv}.
The measured pairwise velocity distributions for all galaxies are plotted with
triangles in the upper panels for MDPL2-\textsc{Sag} (left) and for SDSS-DR7 
(right). The modeled $f(v)$ are shown with the grey solid curve.
For void galaxies, we show measured and modeled $f(v)$ in the bottom panels, 
MDPL2-\textsc{Sag} at left and SDSS-DR7 at right. 
Measurements are shown with diamonds, and with the red solid and blue dashed 
lines the model fits. 
In all cases, the parameter values for $\beta$ and $w$ are shown in the key
figures.

It can be seen in this figure the different shapes of $f(v)$ for all
galaxies and void galaxies, reflecting the smaller redshift distortions observed in
$\xi(s)$ for galaxies within voids (Sec. \ref{sec:xi}).
These differences, observed in both the observations and in the simulated galaxies, 
is quantified by the fitting parameter values $w$ presented in Tab. \ref{tab:fv}. 
While for all galaxies we obtain a value $w\sim 500 \kms$, for void
galaxies the pairwise velocity dispersions range only between $w \sim 50-70 \kms$,
roughly one order of magnitude smaller.

We stress the fact that this method is not particularly sensible to obtain the $\beta$ 
parameter accurately from the derived $f(v)$ distributions, as discussed by
\citet{loveday_gama_2018}.
Furthermore, besides the best fitting pair of values $\beta$ and $w$, we 
have also tested the resulting velocity dispersions derived by fixing the $\beta$
parameter with two fiducial values: for galaxies in voids $\beta=0.00$ (assuming 
completely empty regions), and for all galaxies $\beta=0.50$ (corresponding to 
unbiased tracers in a $\Omega_{\rm m}=0.307$ model).
It should be remarked that regardless the use of either best fitting parameters 
$\beta$ or fixed fiducial values, the derived pairwise velocity dispersion $w$ are 
in agreement within uncertainties. 
This can be appreciated by inspection to Fig. \ref{fig:fv}, where either varying 
or fixed $\beta$ values in the model provide suitable fitting curves to the derived 
$f(v)$.

\begin{figure*}
	\includegraphics[width=\textwidth]{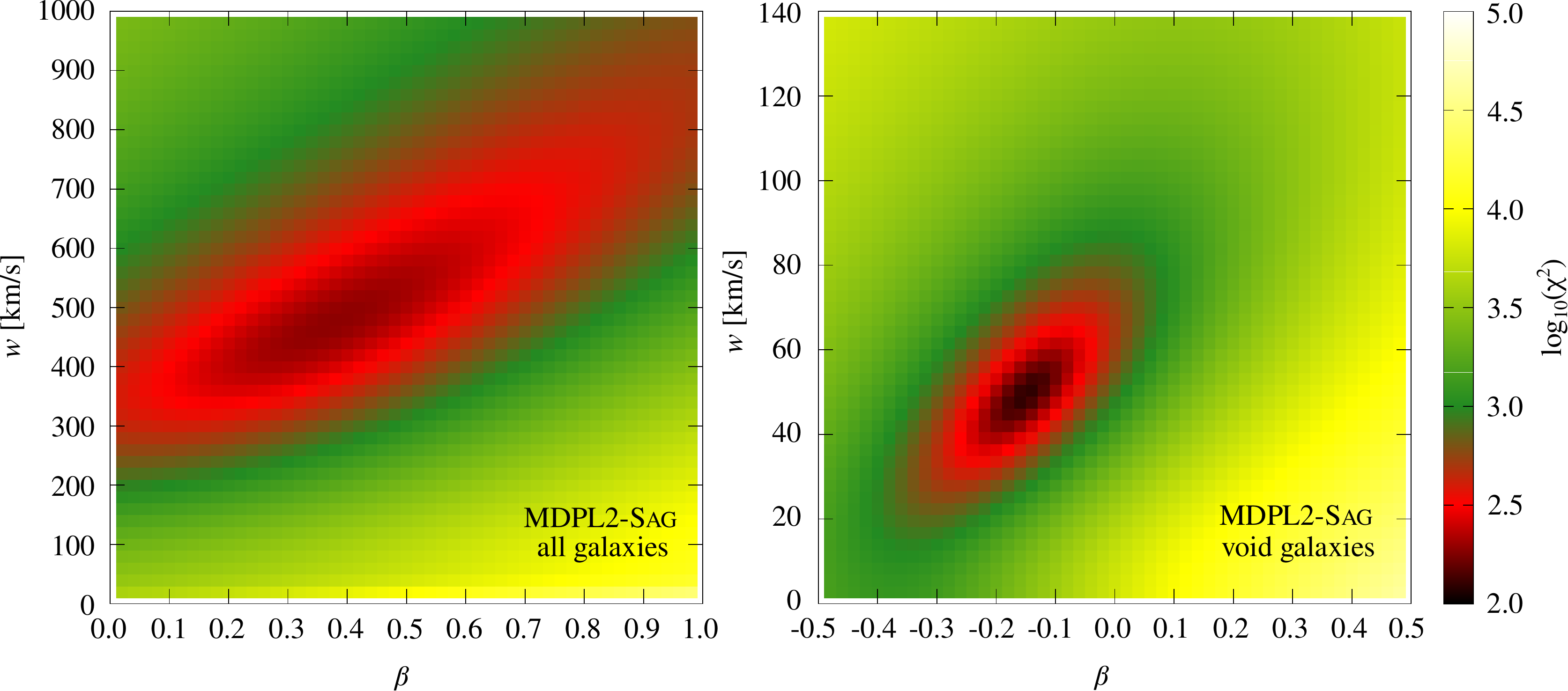}
    \caption{Maps of $\log_{10}(\chi^2)$ resulting from the fitting procedure, 
    described in Sec. \ref{sec:fit} to determine $\beta$ and $w$ for the case 
    of all galaxies (left) and void galaxies (right) in the MDPL2-\textsc{Sag} catalogue.}
    \label{fig:map}
\end{figure*}

The $\beta$ and $w$ degeneracy in the model can be appreciated in
Fig. \ref{fig:map}, where we show the maps of $\log_{10}(\chi^2)$ resulting from 
the fitting procedure to determine these parameters for the case of all galaxies
(left) and void galaxies (right) in the MDPL2-\textsc{Sag} catalogue.
In this figure it can also be seen the large range of allowed $\beta$ values, 
which implies very different cosmological and astrophysical scenarios 
($\Omega_{\rm m}$ and $b$), whilst $w$ values are better constrained.
Nevertheless, this approach is sufficient for our purpose to quantify with a 
simple modeling  the dynamical behavior difference of galaxies in voids with 
respect to the general population.

\begin{figure}
	\includegraphics[width=\columnwidth]{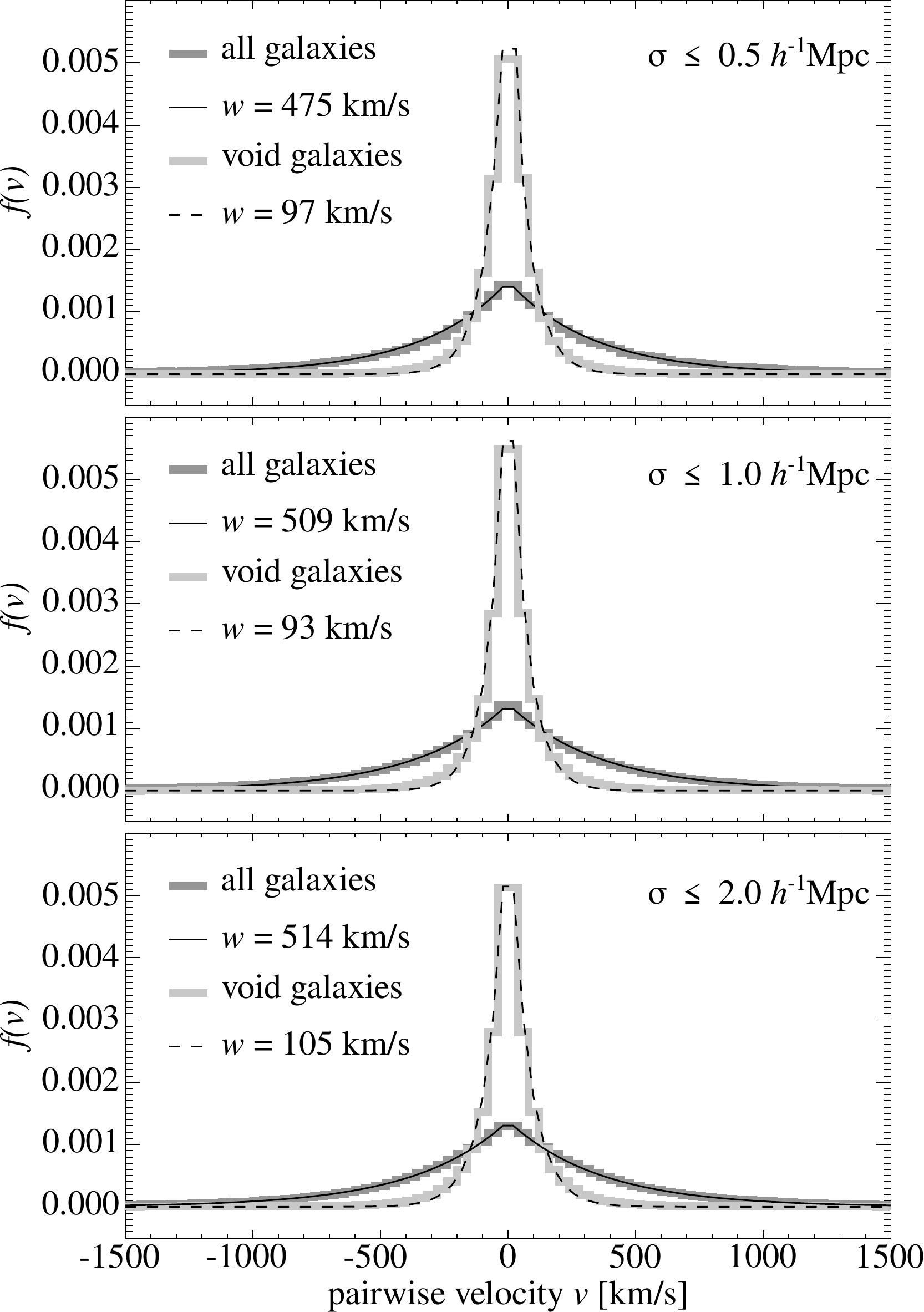}
    \caption{Pairwise velocity distributions measured in MDPL2-\textsc{Sag}
    catalogue. From top to bottom, the panels correspond to measurements of the
    line-of-sight relative velocity distributions for galaxy pairs in projected
    separations $\sigma \le 0.5,~1.0~{\rm and}~2.0\hmpc$, respectively, as 
    indicated in the top right legend of each panel. The distribution for all
    galaxies (void galaxies) are shown in dark (light) grey solid lines. In solid
    and dashed thin black lines are the exponential model fitted to each data sample.}
    \label{fig:fv_mdpl}
\end{figure}

We  have also measured directly the $f(v)$ distributions for MDPL-\textsc{Sag}
galaxies. 
These measures correspond to the line-of-sight relative velocity distributions
for galaxy pairs in projected separations up to $0.5$, $1.0$ and $2.0\hmpc$,
and including all pairs along the line-of-sight direction up to $20\hmpc$.
The results are shown in Fig. \ref{fig:fv_mdpl} where, from top to bottom, we
show the pairwise velocity distributions measured for all galaxies (dark-grey
thick line) and void galaxies (light-grey thick lines) for 
$\sigma~\le~0.5,~1.0~{\rm and}~2.0\hmpc$, respectively.
The thin black lines correspond to exponential models given by Eq.~(\ref{eq:exp})
fitted to the distribution of all galaxies (solid lines) and void galaxies
(dashed lines).
The resulting fitting parameters $w$ given in the key figures are
$w\sim 500\kms$ for all galaxies and $w\sim100\kms$ for void galaxies, which compare
suitably to  those obtained by the methods derived previously in this
Section.

%%%%%%%%%%%%%%%%%%%%%%%%%%%%%%%%%%%%%%%%%%%%%%%%%%%%%%%%%%%%%%%%%%%%%%%%%%%%%%%%%%%%%%%%%%%%%
%%%%%%%%%%%%%%%%%%%%%%%%%%%%%%%%%%%%%%%%%%%%%%%%%%%%%%%%%%%%%%%%%%%%%%%%%%%%%%%%%%%%%%%%%%%%%

\section{Dependence of void internal structure on void environment}
\label{sec:xiRS}

\begin{figure}
	\includegraphics[width=\columnwidth]{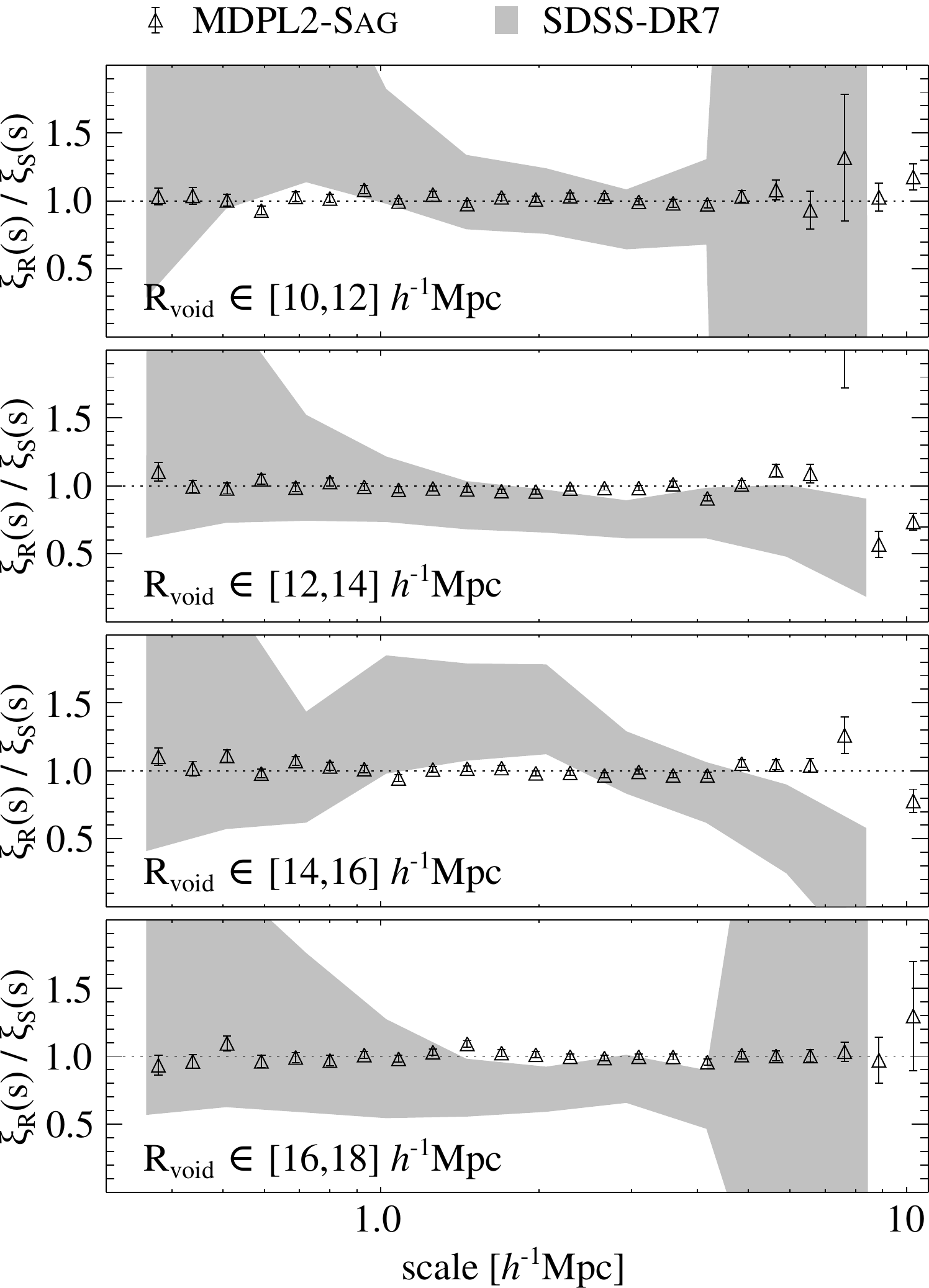}
    \caption{Ratio between the galaxy-galaxy redsfhit-space correlation functions 
    inside R-type voids, $\xi_{\rm R}(s)$, and inside S-type voids, 
    $\xi_{\rm S}(s)$. The panels correspond to different void radii interval, as 
    indicated in the bottom left legends. In all cases, results for MDPL2-\textsc{Sag}
    are shown in black triangles and the results for SDSS-DR7 in grey areas.}
    \label{fig:xi_RS}
\end{figure}

Here we explore the possible dependence of spatial correlations and dynamics of
galaxies within cosmic voids on their large-scale environment.
A useful characterization of the void environment is given by the overdensity
within 2 and 3 void radii
\citep{ceccarelli_clues_2013,paz_clues_2013,ruiz_clues_2015}, which corresponds
to the void-in-void and void-in-cloud early classification by
\citet{sheth_voids_2004}. 
We remark that in spite of their different external environment, both types of 
voids are defined with the same constrains in terms of internal underdensity. 
R-type voids have underdense surroundings implying a continuously rising density 
profile even beyond $R_{\rm void}$, and so their future evolution is dominated by 
expansion.
On the other hand, S-type voids are surrounding by shell-like structures which
generates a collapse forecast of the void itself.  
We argue that by examining the properties of voids internal structure as measured by
galaxy correlations, in both simulations and the observations, can provide
further consistency tests of structure formation in the $\Lambda$CDM scenario.

In Fig. \ref{fig:xi_RS} we show the resulting ratio between the redshift-space
correlation function of void galaxies inside R-type voids, $\xi_{\rm R}(s)$, and inside
S-type voids, $\xi_{\rm S}(s)$.
The black triangles correspond to MDPL2-\textsc{Sag} galaxies and the shaded
regions to the SDSS-DR7 galaxies.
The different panels correspond to different ranges of void radii as indicated
in the key figures.
Remarkably, it can be seen that the internal structure of voids, as measured by
the correlation function, is  unaffected by the external void environment irrespective 
of void radii.
The large uncertainties associated to SDSS-DR7 measurements (given by the shaded areas) 
are due to the low number statistics of voids and galaxy pairs when the complete sample 
is divided both into radii ranges and void types.
We notice however, that within the large uncertainties, the results are consistent with 
the lack of environmental dependence found in MDPL2-\textsc{Sag}.

%%%%%%%%%%%%%%%%%%%%%%%%%%%%%%%%%%%%%%%%%%%%%%%%%%%%%%%%%%%%%%%%%%%%%%%%%%%%%%%%%%%%%%%%%%%%%
%%%%%%%%%%%%%%%%%%%%%%%%%%%%%%%%%%%%%%%%%%%%%%%%%%%%%%%%%%%%%%%%%%%%%%%%%%%%%%%%%%%%%%%%%%%%%
\section{Summary and conclusions}
\label{sec:conclusion}

We have analysed the internal structure of cosmic voids as traced by faint
galaxies using the two-point galaxy-galaxy correlation function. 
The results are compared to the global behaviour of galaxies with similar 
luminosities whose distribution and dynamics are dominated by those galaxies in large groups and clusters.
  
We apply our analysis to a sample of SDSS-DR7 galaxies, and tested our 
results with the semi-analytic catalogue MDPL2-\textsc{Sag}, obtaining 
a general suitable agreement with the observations.

For galaxies in voids, we  obtain a similar shape for redshift-space and
real-space correlation functions except for a nearly constant normalization
factor which indicates small departures due to peculiar velocities of growing
structure in voids. 
We find that correlations are remarkably linear in voids as compared with
the general population which exhibit large departures due to the velocity
dispersions of galaxies in clusters, absent in cosmic voids.
 
We have also computed the 2D correlation function $\xi(\sigma,\pi)$, where the 
large redshift-space distortions due to galaxy velocity dispersion in virialized systems is
clearly absent for galaxies within voids compared to the global population.

From $\xi(\sigma,\pi)$, we infer the pairwise velocity distributions $f(v)$ of 
galaxies in cosmic voids and elsewhere for MDPL2-\textsc{Sag} and SDSS-DR7. 
For both observations and simulations, we find these distributions consistent
with an exponential model with a pairwise velocity dispersion $w \sim 50-70 \kms$, a
figure significantly smaller than the derived global values of $w \sim 500
\kms$. 
We recall the consistency of model and observational determinations which 
provide further support to the standard $\Lambda$CDM scenario of structure formation 
in a less explored regime of low-luminosity galaxies residing in global underdense 
regions.

For MDPL2-\textsc{Sag} galaxies, in addition to the derivation of $f(v)$ from 
$\xi(\sigma,\pi)$, we have computed directly for three different ranges of 
projected distance $\sigma \le 0.5,~1.0,~{\rm and}~2.0\hmpc$. We find that  
in all cases the pairwise velocity distribution can be modeled by 
exponential functions with best fitting velocity dispersion parameters 
$w\sim 500\kms$ for all galaxies and $w\sim 100\kms$ for void galaxies, entirely 
consistent with the previous analysis.

Finally, we have tested the potential influence of the large-scale structure 
surrounding cosmic voids on the clustering of galaxies inside voids.
We have compared the resulting $\xi(s)$ for galaxies in void-in-void (R-type voids) 
and void-in-cloud (S-type voids) \citep{sheth_voids_2004,ceccarelli_clues_2013},
for several void radius intervals, finding the same clustering amplitude of faint 
galaxies independent of the void environment. Namely, we find no influence of the 
surrounding structure beyond $R_{\rm void}$ on the internal void structure as traced 
by faint galaxies.

Our results show the potentiality of exploring the structure of cosmic voids
through the distribution of faint galaxies.
These void galaxies are not largely affected by strong non-linearities nor 
by gas dynamical effects, and their distribution is not influenced by  the 
large-scale environment of voids. 
For these reasons, we argue that galaxies in voids may provide suitable samples 
to analyse possible subtle differences between the observations and current models of 
structure formation. 
These analysis may be extended in upcoming future galaxy surveys which can 
provide large datasets capable of extending our understanding of the spatial
distribution and the dynamics of luminous and dark matter. 

\section*{acknowledgements}

We kindly thank to the referee Dr. Yan-Chuan Cai for his very useful comments 
and suggestions that helped to improve this paper.
The authors also thank Nelson D. Padilla for useful discussions and comments.

This work was partially supported by the Consejo Nacional de Investigaciones
Cient\'{\i}ficas y T\'ecnicas (CONICET), and the Secretar\'{\i}a de Ciencia y
Tecnolog\'{\i}a (SeCyT), Universidad Nacional de C\'ordoba, Argentina.

This research has made use of NASA's Astrophysics Data System. 

%MDPL acknowledgement
The \textsc{CosmoSim} database used in this paper is a service by the
Leibniz-Institute for Astrophysics Potsdam (AIP). The authors gratefully
acknowledge the Gauss Centre for Supercomputing e.V. (www.gauss-centre.eu) and
the Partnership for Advanced Supercomputing in Europe (PRACE, www.prace-ri.eu)
for funding the \textsc{MultiDark} simulation project by providing computing
time on the GCS Supercomputer SuperMUC at Leibniz Supercomputing Centre (LRZ,
www.lrz.de).

%SDSS acknowledgement
Funding for the SDSS and SDSS-II has been provided by the Alfred P. Sloan
Foundation, the Participating Institutions, the National Science Foundation,
the U.S. Department of Energy, the National Aeronautics and Space
Administration, the Japanese Monbukagakusho, the Max Planck Society, and the
Higher Education Funding Council for England. The SDSS Web Site is
http://www.sdss.org/. The SDSS is managed by the Astrophysical Research
Consortium for the Participating Institutions. The Participating Institutions
are the American Museum of Natural History, Astrophysical Institute Potsdam,
University of Basel, University of Cambridge, Case Western Reserve University,
University of Chicago, Drexel University, Fermilab, the Institute for Advanced
Study, the Japan Participation Group, Johns Hopkins University, the Joint
Institute for Nuclear Astrophysics, the Kavli Institute for Particle
Astrophysics and Cosmology, the Korean Scientist Group, the Chinese Academy of
Sciences (LAMOST), Los Alamos National Laboratory, the Max-Planck-Institute for
Astronomy (MPIA), the Max-Planck-Institute for Astrophysics (MPA), New Mexico
State University, Ohio State University, University of Pittsburgh, University
of Portsmouth, Princeton University, the United States Naval Observatory, and
the University of Washington.

%%%%%%%%%%%%%%%%%%%% REFERENCES %%%%%%%%%%%%%%%%%%
\bibliographystyle{mnras}
\bibliography{references}
\bsp  % typesetting comment
\label{lastpage}

\end{document}